\documentclass[12pt,a4paper]{cernart}
\usepackage{epsfig}
\usepackage{amssymb}
\usepackage{cite}
\usepackage{colordvi}

\def\lapproxeq{\lower .7ex\hbox{$\stackrel{\textstyle
<}{\sim}$}}
\def\gapproxeq{\lower .7ex\hbox{$\stackrel{\textstyle
>}{\sim}$}}
\newcommand{\GeV}{\mathrm{GeV}}

\def\Qs{$Q^2$}
\def\rn{$\rho^{0}$}			
\def\pip{$\pi^{+}$}			
\def\pin{$\pi^{-}$}			
\def\mpipi{$m_{\pi\pi}$}		
\def\Emi{$E_{miss}$}			
\def\xBj{$x$}            
\def\pts{$p_{t}^{2}$}			
\def\Aor{$A_{1}^{\rho}$}		
\def\ALL{$A_{LL}$}			

\begin {document}
\dimen\footins=\textheight

\begin{titlepage}
\docnum{\vbox{CERN--PH--EP/2007--009\\
(revised author list)}}
\date{2 April 2007}

\vspace{1cm}

\begin{center}
{\LARGE {\bf Double spin asymmetry in exclusive $\rho^0$ muoproduction at COMPASS
              }}
\vspace*{0.5cm}
\end{center}

\vspace{2cm}

\begin{abstract}
The longitudinal double spin asymmetry $A_1^{\rho}$ for exclusive
leptoproduction of $\rho ^0$ mesons, $\mu  +  N  \rightarrow 
\mu  +  N  +  \rho$, is studied using the COMPASS 2002 and 2003 data.
The measured reaction is incoherent exclusive $\rho^{0}$ production on
polarised deuterons.
The \Qs\ and \xBj\ dependence of $A_1^{\rho}$ is presented in a wide kinematical range
$3 \cdot 10^{-3} < Q^2 < 7~(\GeV/c)^2$ and $5 \cdot 10^{-5} < x < 0.05$.
The presented results are the first measurements of $A_1^{\rho}$ at small
$Q^2$ ($Q^2 < 0.1~(\GeV/c)^2$) and small \xBj\ ($x < 3 \cdot 10^{-3}$).
The asymmetry is in general compatible with zero in the whole kinematical
range.

\vfill
\submitted{(to be submitted to Eur.\ Phys.\ J. C)}
\end{abstract}
\begin{Authlist}
{\large  The COMPASS Collaboration}\\[\baselineskip]
%
%
M.~Alekseev\Iref{turin},
V.Yu.~Alexakhin\Iref{dubna},
Yu.~Alexandrov\Iref{moscowlpi},
G.D.~Alexeev\Iref{dubna},
A.~Amoroso\Iref{turin},
A.~Arbuzov\Iref{dubna},
B.~Bade\l ek\Iref{warsaw},
F.~Balestra\Iref{turin},
J.~Ball\Iref{saclay},
G.~Baum\Iref{bielefeld}
% ...
\end{Authlist}
%
%
\Instfoot{bielefeld}{ Universit\"at Bielefeld, Fakult\"at f\"ur Physik, 33501 Bielefeld, Germany\Aref{e}}
\Instfoot{bochum}{ Universit\"at Bochum, Institut f\"ur Experimentalphysik, 44780 Bochum, Germany\Aref{e}}
\Instfoot{bonniskp}{ Universit\"at Bonn, Helmholtz-Institut f\"ur  Strahlen- und Kernphysik, 53115 Bonn, Germany\Aref{e}}
\Instfoot{bonnpi}{ Universit\"at Bonn, Physikalisches Institut, 53115 Bonn, Germany\Aref{e}}
\Instfoot{brno}{Institute of Scientific Instruments, AS CR, 61264 Brno, Czech Republic\Aref{f}}
\Instfoot{burdwan}{ Burdwan University, Burdwan 713104, India\Aref{h}}
\Instfoot{calcutta}{ Matrivani Institute of Experimental Research \& Education, Calcutta-700 030, India\Aref{i}}
\Instfoot{dubna}{ Joint Institute for Nuclear Research, 141980 Dubna, Moscow region, Russia}
\Instfoot{erlangen}{ Universit\"at Erlangen--N\"urnberg, Physikalisches Institut, 91054 Erlangen, Germany\Aref{e}}
\Instfoot{freiburg}{ Universit\"at Freiburg, Physikalisches Institut, 79104 Freiburg, Germany\Aref{e}}
\Instfoot{cern}{ CERN, 1211 Geneva 23, Switzerland}
\Instfoot{heidelberg}{ Universit\"at Heidelberg, Physikalisches Institut,  69120 Heidelberg, Germany\Aref{e}}
\Instfoot{helsinki}{ Helsinki University of Technology, Low Temperature Laboratory, 02015 HUT, Finland  and University of Helsinki, Helsinki Institute of  Physics, 00014 Helsinki, Finland}
\Instfoot{liberec}{Technical University in Liberec, 46117 Liberec, Czech Republic\Aref{f}}
\Instfoot{lisbon}{ LIP, 1000-149 Lisbon, Portugal\Aref{g}}
\Instfoot{mainz}{ Universit\"at Mainz, Institut f\"ur Kernphysik, 55099 Mainz, Germany\Aref{e}}
\Instfoot{miyazaki}{University of Miyazaki, Miyazaki 889-2192, Japan\Aref{j}}
\Instfoot{moscowlpi}{Lebedev Physical Institute, 119991 Moscow, Russia}
\Instfoot{munichlmu}{Ludwig-Maximilians-Universit\"at M\"unchen, Department f\"ur Physik, 80799 Munich, Germany\Aref{e}}
\Instfoot{munichtu}{Technische Universit\"at M\"unchen, Physik Department, 85748 Garching, Germany\Aref{e}}
\Instfoot{nagoya}{Nagoya University, 464 Nagoya, Japan\Aref{j}}
\Instfoot{praguecu}{Charles University, Faculty of Mathematics and Physics, 18000 Prague, Czech Republic\Aref{f}}
\Instfoot{praguectu}{Czech Technical University in Prague, 16636 Prague, Czech Republic\Aref{f}}
\Instfoot{protvino}{ State Research Center of the Russian Federation, Institute for High Energy Physics, 142281 Protvino, Russia}
\Instfoot{saclay}{ CEA DAPNIA/SPhN Saclay, 91191 Gif-sur-Yvette, France}
\Instfoot{telaviv}{ Tel Aviv University, School of Physics and Astronomy, 
              69978 Tel Aviv, Israel\Aref{k}}
\Instfoot{triestictp}{ ICTP--INFN MLab Laboratory, 34014 Trieste, Italy}
\Instfoot{triest}{ INFN Trieste and University of Trieste, Department of Physics, 34127 Trieste, Italy}
\Instfoot{turin}{ INFN Turin and University of Turin, Physics Department, 10125 Turin, Italy}
\Instfoot{warsaw}{ So{\l}tan Institute for Nuclear Studies and Warsaw University, 00-681 Warsaw, Poland\Aref{l} }
\Instfoot{warsawtu}{ Warsaw University of Technology, Institute of Radioelectronics, 00-665 Warsaw, Poland\Aref{m} }
\Instfoot{yamagata}{ Yamagata University, Yamagata, 992-8510 Japan\Aref{j} }
\Anotfoot{a}{Also at IST, Universidade T\'ecnica de Lisboa, Lisbon, Portugal}
\Anotfoot{b}{Also at University of East Piedmont, 15100 Alessandria, Italy}
\Anotfoot{c}{On leave of absence from JINR Dubna}               
\Anotfoot{d}{Also at Chubu University, Kasugai, Aichi, 487-8501 Japan}
\Anotfoot{e}{Supported by the German Bundesministerium f\"ur Bildung und Forschung}
\Anotfoot{f}{Suppported by Czech Republic MEYS grants ME492 and LA242}
\Anotfoot{g}{Supported by the Portuguese FCT - Funda\c{c}\~ao para
               a Ci\^encia e Tecnologia grants POCTI/FNU/49501/2002 and POCTI/FNU/50192/2003}
\Anotfoot{h}{Supported by DST-FIST II grants, Govt. of India}
\Anotfoot{i}{Supported by  the Shailabala Biswas Education Trust}
\Anotfoot{j}{Supported by the Ministry of Education, Culture, Sports,
               Science and Technology, Japan; Daikou Foundation and Yamada Foundation}
\Anotfoot{k}{Supported by the Israel Science Foundation, founded by the Israel Academy of Sciences and Humanities}
\Anotfoot{l}{Supported by KBN grant nr 621/E-78/SPUB-M/CERN/P-03/DZ 298 2000,
               nr 621/E-78/SPB/CERN/P-03/DWM 576/2003-2006, and by MNII reasearch funds for 2005--2007}
\Anotfoot{m}{Supported by  KBN grant nr 134/E-365/SPUB-M/CERN/P-03/DZ299/2000}

\vfill

\end{titlepage}

\newpage
~
\newpage
\newpage
\section{Introduction}
 \label{Sec_intro}

  In this paper we present results on the longitudinal double spin
asymmetry $A_1^{\rho }$ for exclusive incoherent  $\rho ^0$ production in the
scattering of high energy muons on nucleons.
The experiment was carried out at CERN by the COMPASS collaboration
using the 160~GeV muon beam and the large $^{6}$LiD polarised target.

The studied reaction is 
\begin{equation} 
 \mu  +  N  \rightarrow  \mu^{\prime}  +  \rho ^0  +  N^{\prime},       \label{murho}
\end{equation}
where $N$ is a quasi-free nucleon from the polarised deuterons.
The reaction (\ref{murho}) can be described in terms of the virtual photoproduction
process
\begin{equation}
 \gamma^{\ast}  +  N  \rightarrow  \rho^0  +  N^{\prime}.   \label{phorho} 
\end{equation}
The reaction (\ref{phorho}) can be regarded as a fluctuation of the
virtual photon into a quark-antiquark pair (in partonic language),
or an off-shell vector meson (in Vector Meson Dominance model), which then scatters
off the target nucleon resulting in the production of an on-shell vector meson.
At high energies this is predominantly a diffractive process and plays
an important role in the investigation of Pomeron exchange and its
interpretation in terms of multiple gluon exchange.

Most of the presently available information on the spin structure 
of reaction (\ref{phorho})
stems from the $\rho ^0$ spin density matrix elements,
which are obtained from the analysis of angular distributions
of  $\rho ^0$ production and decay \cite{Schilling73}. Experimental results
on $\rho ^0$ spin density matrix elements 
come from various experiments \cite{NMC,E665-1,ZEUS,H1,HERMES00} including
the preliminary results from COMPASS \cite{sandacz}.
 
The emerging picture of the spin structure of the considered process 
is the following. At low photon virtuality $Q^2$ the cross section by transverse virtual photons
$\sigma _T$
dominates, while the relative contribution of the cross section by longitudinal
photons $\sigma _L$ 
rapidly increases with $Q^2$. At $Q^2$ of about 2~(GeV/{\it c})$^2$ both
components become comparable and at a larger $Q^2$ the contribution of
$\sigma _L$ becomes dominant and continues to grow, although
at lower rate than at low $Q^2$. 
Approximately, the so called $s$-channel helicity
conservation (SCHC) is valid, i.e.\ the helicity of the vector meson is the same
as the helicity of the parent virtual photon. The data indicate that 
the process can be described approximately by the 
exchange in the $t$-channel of an object with natural 
parity $P$.
Small deviations from SCHC are observed, also at the highest energies, whose
origin is still to be understood. 
An interesting suggestion was made in Ref.~\cite{igivanov} that at high energies
the magnitudes of various helicity amplitudes for the reaction (\ref{phorho})
may shed a light on the spin-orbital momentum structure of the vector meson. 

A complementary information can be obtained from measurements of the double
spin cross section asymmetry, when the information on both the beam
and target polarisation is used.
The asymmetry is defined as
\begin{equation}
 A_{1}^{\rho}  = 
   \frac{
    \sigma_{1/2}  -  \sigma_{3/2}
   }{
    \sigma_{1/2}  +  \sigma_{3/2}}
   ,
   \label{A1def}
\end{equation}
where $\sigma_{1/2  (3/2)}$ stands for the cross sections of the reaction (\ref{phorho})
and the subscripts denote the total virtual photon--nucleon angular momentum
component along the virtual photon
direction.
In the following we will also use the asymmetry
\ALL\
which is defined for reaction (\ref{murho}) as the asymmetry of
muon--nucleon cross sections for antiparallel and parallel beam and
target longitudinal spin orientations.

In the Regge approach \cite{manaenkov} the longitudinal double spin asymmetry
$A_1^{\rho }$ can arise due 
to the interference of amplitudes for exchange in the $t$-channel of Reggeons with natural parity
(Pomeron, $\rho $, $\omega $, $f$, $A_2$ ) with amplitudes for Reggeons with 
unnatural parity 
($\pi , A_1$).
No significant asymmetry is expected 
when only a non-perturbative 
Pomeron is exchanged because it has small spin-dependent couplings as found from 
hadron-nucleon data for cross sections and polarisations.

Similarly, in the approach of Fraas \cite{fraas76}, 
assuming approximate validity of 
SCHC, the spin asymmetry $A_1^{\rho }$ arises from the interference between
parts of the helicity amplitudes for transverse photons corresponding
to the natural and unnatural parity exchanges
in the $t$ channel. While a measurable asymmetry can arise even from a small
contribution of the unnatural parity exchange, the latter may remain
unmeasurable in the cross sections.
A significant unnatural-parity contribution may indicate an exchange 
of certain Reggeons like $\pi$, $A_{1}$ or in
partonic terms an exchange of $q\bar{q}$ pairs.

In the same reference a theoretical prediction for $A_1^{\rho}$ was
presented, which is based on the description of forward exclusive $\rho^{0}$
leptoproduction and inclusive inelastic lepton-nucleon
scattering by the off-diagonal Generalised Vector Meson Dominance (GVMD) model,
applied to the case of polarised lepton--nucleon scattering.
At the values of Bjorken variable $x < 0.2$, with additional assumptions
\cite{HERMES01}, \Aor\ can be related 
to the $A_{1}$ asymmetry for inclusive inelastic lepton scattering at the same
\xBj\ as
\begin{equation}
 A_1^{\rho }  =  \frac{2 A_1}{1  +  (A_1)^2} .    \label{A1-Fraas}
\end{equation} 
This prediction is consistent with the HERMES results for both the proton
and deuteron targets, although with rather large errors.

In perturbative QCD, there exists a general proof of factorisation \cite{fact}
for exclusive vector meson production by longitudinal photons. 
It allows a decomposition of the full amplitude for reaction (\ref{phorho})
into three components:
a hard scattering amplitude for the exchange of quarks or gluons,
a distribution amplitude for the meson and 
the non-perturbative description of the target nucleon in terms of
the generalised parton distributions (GPDs), which are related to the internal
structure
of the nucleon.
No similar proof of factorisation exists for transverse virtual photons, and as a consequence
the interpretation of $A_{1}^{\rho}$ in perturbative QCD is not possible
at leading twist. However, a model including higher twist effects proposed
by Martin et al. \cite{mrt} describes the behaviour of both $\sigma_{L}$
as well as of $\sigma _T$ reasonably well. An extension of this model by Ryskin
\cite{Ryskin} for the
spin dependent cross sections allows to relate
\Aor\ to the spin dependent GPDs of gluons and quarks in the nucleon.
The applicability of this model is limited to the range $Q^2 \geq 4~(\GeV/c)^2$.
More recently another pQCD-inspired model involving GPDs has been proposed by Goloskokov and Kroll 
\cite{krgo,gokr}. The non-leading twist asymmetry $A_{LL}$ results from the 
interference between the dominant GPD $H_g$ and the helicity-dependent GPD
$\tilde{H}_g$. The asymmetry is estimated to be of the order 
$ k_T^2  \tilde{H}_g / (Q^2  H_g  )$,
where $k_T$ is the transverse momentum of the quark and the antiquark.

Up to now little experimental information has been available on the
double spin asymmetries for exclusive leptoproduction of vector mesons. 
The first observation of a non-zero asymmetry $A_1^{\rho }$ in polarised
electron--proton deep-inelastic scattering was reported by the HERMES experiment 
\cite{HERMES01}.
In the deep inelastic region $(0.8 < Q^2 < 3~(\GeV/c)^2)$
the measured asymmetry is equal to 0.23 $\pm$ 0.14 (stat) $\pm$ 0.02 (syst)
\cite{HERMES03},
with little dependence on the kinematical variables. In contrast, for the `quasi-real photoproduction' data, with 
$\langle Q^2\rangle = 0.13~(\GeV/c)^2$, the asymmetry for the proton target is consistent with zero.
On the other hand the measured asymmetry $A_1^{\rho }$ for the polarised deuteron target and the asymmetry $A_1^{\phi }$ for exclusive production of $\phi $
meson 
either on polarised protons or deuterons
 are consistent with zero both in the deep inelastic and in the 
quasi-real photoproduction regions
 \cite{HERMES03}.

The HERMES result indicating a non-zero $A_1^{\rho }$ for the proton
target differs from the unpublished result of similar measurements by 
the SMC experiment \cite{ATrip-1} at comparable values of \Qs\ but at
about three times higher values of the photon-nucleon centre of mass energy $W$, i.e.\ at smaller \xBj. The SMC measurements of 
\ALL\ 
in several bins of $Q^2$ are consistent with zero
for both proton and deuteron targets. 

\section{The experimental set-up}
 \label{Sec_exper}

The experiment \cite{setup} was performed with
the high intensity positive muon beam from the CERN M2 beam line.
The $\mu ^{+}$ beam intensity is $2 \cdot 10^8$ per spill of 4.8 s with a cycle time of
16.8~s. The average beam energy is 160~GeV and the momentum spread is $\sigma _{p}/p = 0.05$.
The momentum of each beam muon is measured upstream of the experimental area in a beam
momentum station consisting of several planes of scintillator strips or scintillating fibres
with a dipole magnet in between. The precision of the momentum determination is typically
$\Delta p/p \leq 0.003$. The $\mu ^{+}$ beam is naturally polarised by the weak decays 
of the parent
hadrons. The polarisation of the muon varies with its energy and the 
average polarisation is $-0.76$.

The beam traverses the two cells of the polarised target, 
each 60~cm long, 3~cm in diameter and separated by 10~cm, which are placed one after the other. 
The target cells are filled with $^{6}$LiD which is used as polarised deuteron target material
and is longitudinally polarised by dynamic nuclear polarisation (DNP).
The two cells are polarised in opposite directions so that data from both spin directions
are recorded at the same time. The typical values of polarisation are about 0.50.
A mixture of liquid $^3$He
and $^4$He, used to refrigerate the target, and a small amount of heavier nuclei are also
present in the target.  
The spin directions in the two target cells are reversed every 8 hours by rotating
the direction of the magnetic field in the target. 
In this way fluxes and acceptances cancel in the calculation of spin asymmetries, provided that
the ratio of acceptances of the two cells remains unchanged after the reversal. 

The COMPASS spectrometer is designed to reconstruct the scattered muons
and the produced hadrons in wide momentum and angular ranges.
It is divided in two stages with two dipole magnets, SM1 and SM2. The first magnet, 
SM1, accepts charged particles of momenta larger than 0.4~GeV/{\it c}, and the second one, SM2, those
larger than 4~GeV/{\it c}. The angular acceptance of the spectrometer is limited by
the aperture of the polarised target magnet. For the upstream
end of the target it is $\pm 70$~mrad.

To match the expected particle flux at various locations in the spectrometer, COMPASS
uses various tracking detectors. Small-angle tracking is provided by stations of
scintillating fibres, silicon detectors, micromesh gaseous chambers and gas electron
multiplier chambers. Large-angle tracking devices are multiwire proportional chambers,
drift chambers and straw detectors. Muons are identified in large-area mini drift 
tubes and drift tubes placed downstream of hadron absorbers. Hadrons are detected by
two large iron-scintillator sampling calorimeters installed in front of the absorbers
and shielded to avoid electromagnetic contamination.
The identification of charged particles is possible with a RICH detector, although
in this paper we have not utilised the information from the RICH.

The data recording system is activated by various triggers indicating the presence
of a scattered muon and/or an energy deposited by hadrons
in the calorimeters. In addition to the inclusive trigger, in which the scattered muon
is identified by coincidence signals in the trigger hodoscopes, several semi-inclusive
triggers were used. They select events fulfilling the requirement to detect
the scattered muon together with the energy deposited in the hadron calorimeters exceeding
a given threshold. In 2003 the acceptance was further extended towards high $Q^2$
values by the addition of a standalone calorimetric trigger in which no condition 
is set for the scattered muon. 
The COMPASS trigger system allows us to cover a wide range of  $Q^2$, from quasi-real
photoproduction to deep inelastic interactions.
 
A more detailed description of the COMPASS apparatus can be found in Ref.~\cite{setup}

\section{Event sample}
 \label{Sec_sample}

For the present analysis the whole data sample taken in 2002 and 2003 with the 
longitudinally polarised target is used. For an event to be accepted for further analysis it is required to originate in the target, have a reconstructed 
beam track, a
scattered  muon track, and only two additional tracks 
of oppositely charged hadrons associated to the primary vertex. 
The fluxes of beam muons passing through each target cell are equalised using
appropriate cuts on the position and angle of the beam tracks.

The charged
pion mass hypothesis is assigned to each hadron track and the invariant mass of two
pions, $m_{\pi \pi}$, calculated. A cut on the invariant mass of two pions, $0.5 < m_{\pi \pi} < 1~\GeV/c^2$, is applied to select
the $\rho ^0$.
As slow recoil target particles are not detected, in order to select
exclusive events we use the cut on the missing energy, $-2.5 < E_{miss} < 2.5~\GeV$, and on the transverse momentum of $\rho^0$ with respect to the direction of
virtual photon, $p_t^2 < 0.5~(\GeV/c)^2$. Here $E_{miss} = (M^{2}_{X} - M^{2}_{p})/2M_{p}$, where $M_X$ is the missing mass of the unobserved recoiling
system and $M_p$ is the proton mass.
Coherent interactions on the
target nuclei are removed by a cut $p_t^2 > 0.15~(\rm{GeV}/c)^2$. 
To avoid large corrections for acceptance and misidentification of
events, additional cuts 
$\nu > 30~\GeV$ and $E_{\mu'} > 20~\GeV$ are applied.

\begin{figure}[t]
 \begin{center}
 \epsfig{figure=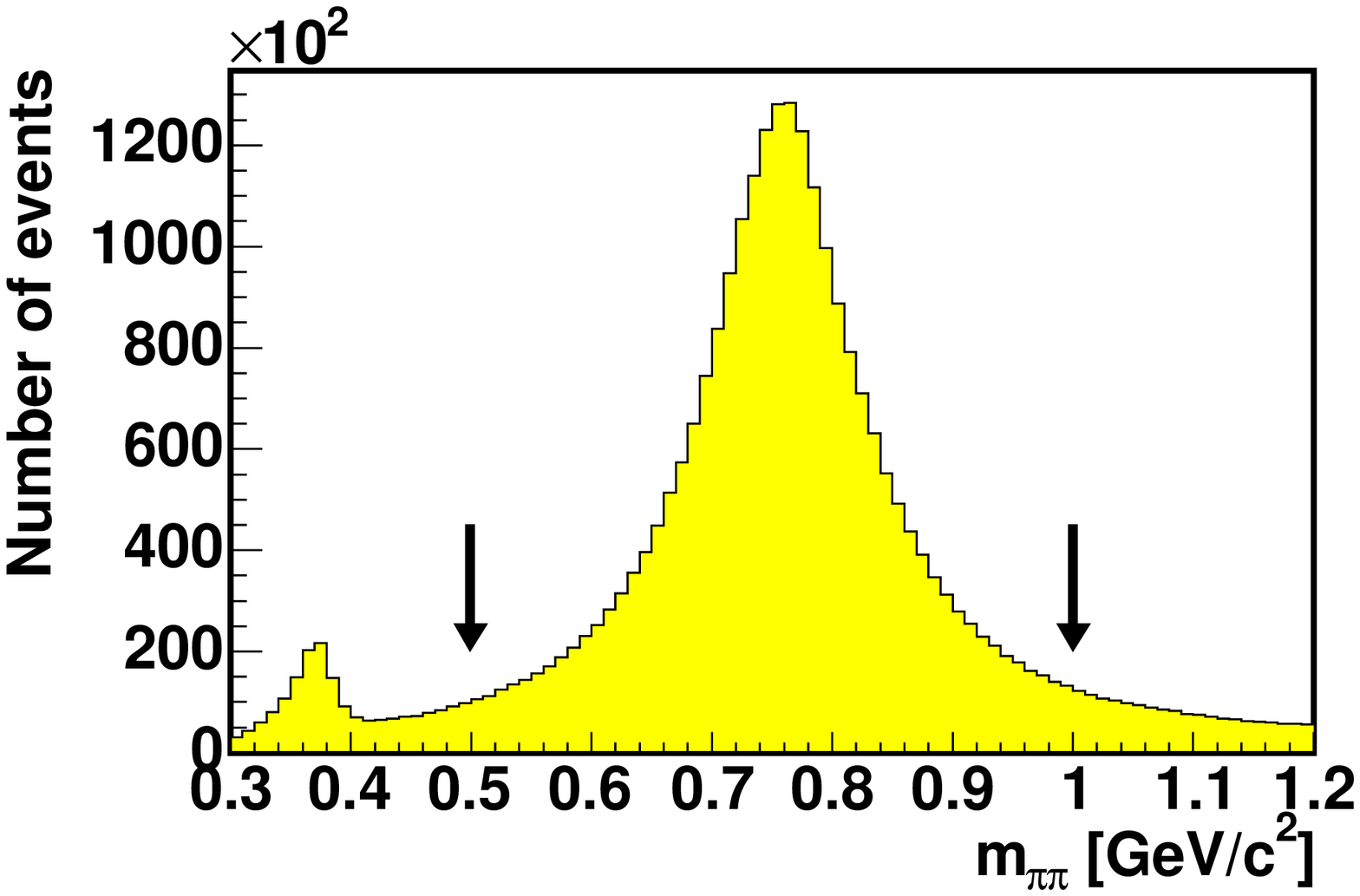,height=5cm,width=7.5cm}
 \epsfig{figure=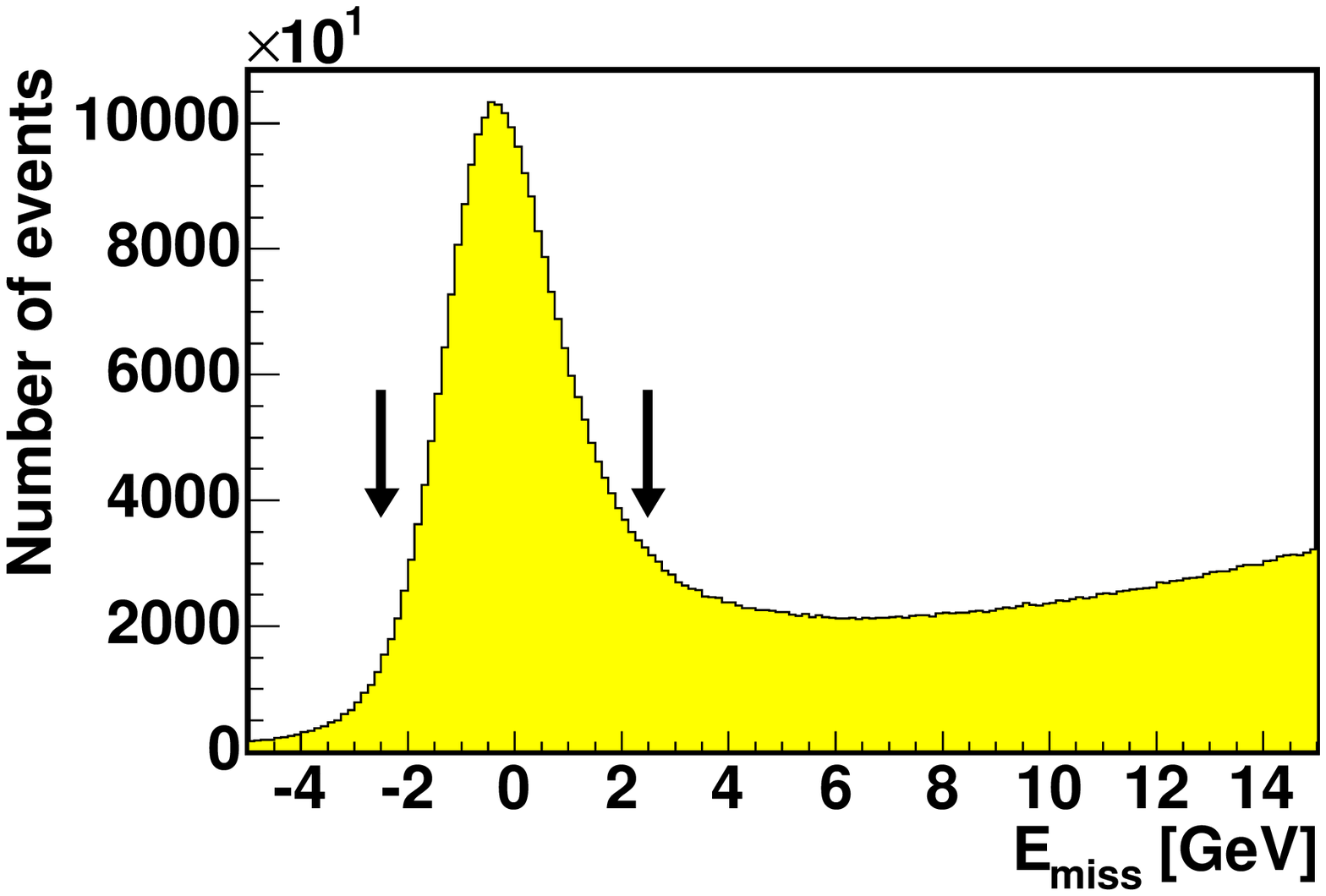,height=5cm,width=7.5cm}
 \epsfig{figure=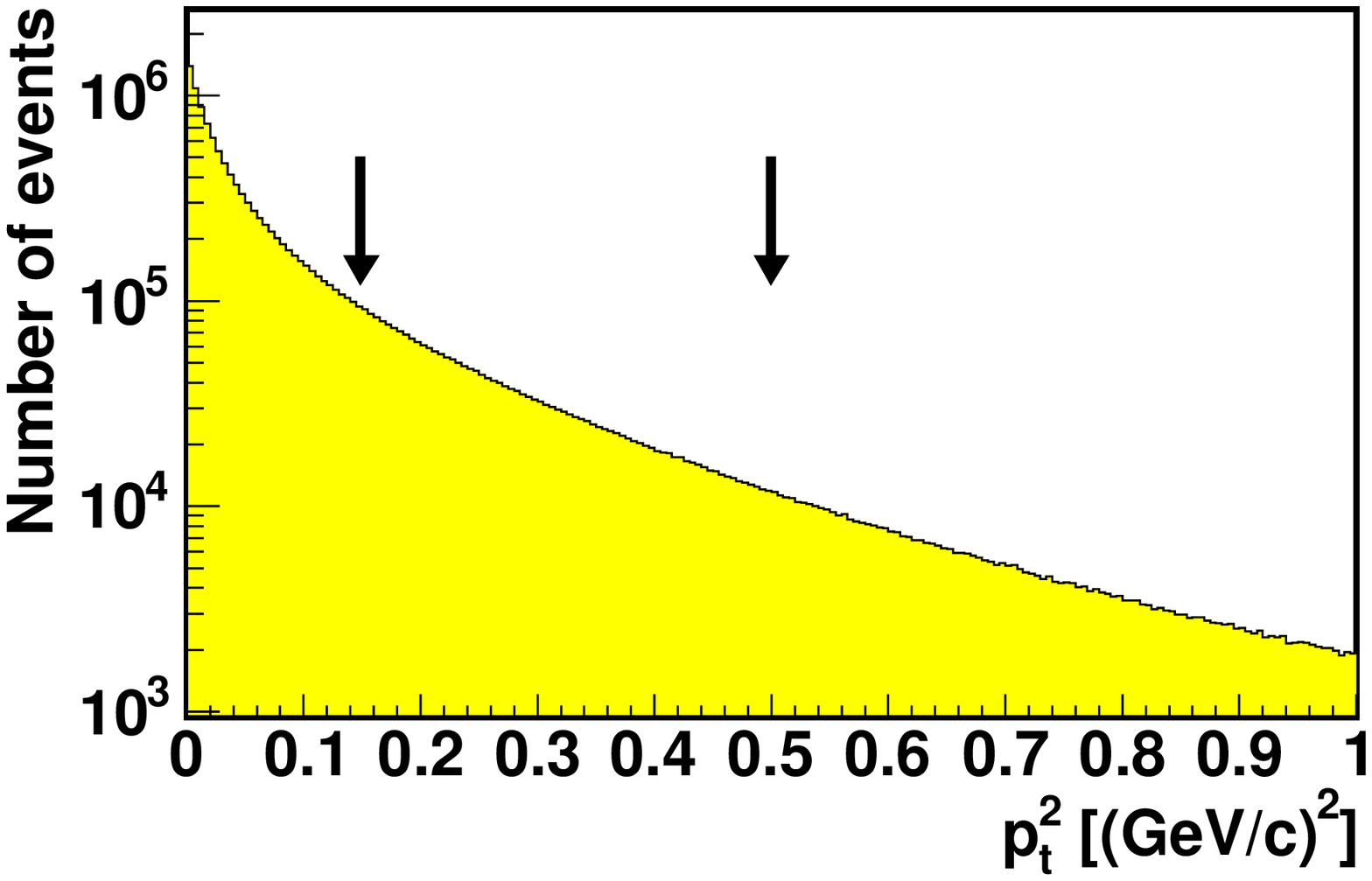,height=5cm,width=7.5cm}
 \caption{Distributions of \mpipi\ (top left), \Emi\ (top right) 
and \pts\ (bottom) for the exclusive sample. 
The arrows show cuts imposed on each variable to define the final sample.}
 \label{hadplots}
 \end{center}
\end{figure}

The distributions of $m_{\pi \pi}$, $E_{miss}$ and $p_t^2$ are shown in 
Fig.~\ref{hadplots}. Each plot is obtained applying all cuts except
those corresponding to the displayed variable. 
On the left top panel of Fig.~\ref{hadplots} a clear peak of the \rn\ resonance, centred at 770~MeV/$c^2$, is visible on the top of the small contribution of
background of the non-resonant \pip\pin\ pairs. Also the skewing of the resonance peak towards smaller values
of \mpipi, due to an interference with the non-resonant background, is noticeable. A small bump below 0.4~GeV/$c^2$
is due to assignment of the charged pion mass to the kaons from decays of $\phi$ mesons.
The mass cuts eliminate the non-resonant background outside of the \rn\ peak, as well as the contribution of $\phi$
mesons.

On the right top panel of the figure the peak at $E_{miss} \approx 0$ is the signal of 
exclusive \rn\ production. The width of the peak, $\sigma \approx 1.1~\GeV$,
is due to the spectrometer resolution. Non-exclusive events, where in addition to the recoil nucleon other undetected hadrons are produced, appear at $E_{miss} > 0$.
Due to the finite resolution, however, they are not resolved from the exclusive peak.
This background
consists of two components: the double-diffractive events where additionally to \rn\ an excited nucleon state is produced in the
nucleon vertex of reaction (\ref{phorho}), and events with semi-inclusive \rn\ production, in which other
hadrons are produced but escape detection. 

The $p_t^2$ distribution shown on the bottom panel of the figure
indicates a contribution from coherent production on target 
nuclei at small $p_t^2$ values. A three-exponential fit to this distribution 
was performed, which indicates also 
a contribution of non-exclusive background increasing
with $p_t^2$.
Therefore to select the sample of exclusive incoherent \rn\ production, 
the aforementioned $p_t^2$ cuts, indicated by arrows, were applied. 

\begin{figure}[t]
 \begin{center}
 \epsfig{figure=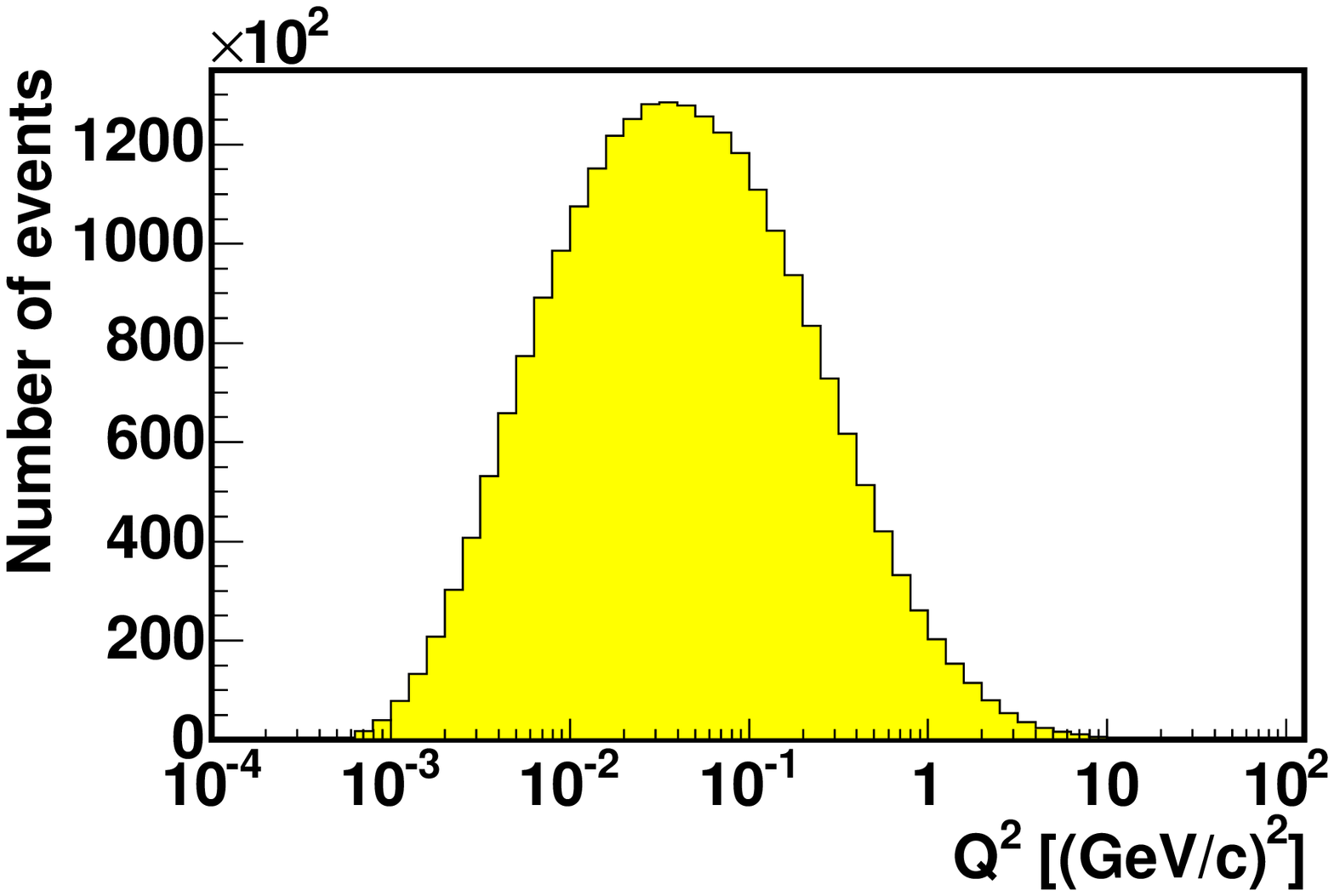,height=5cm,width=7.5cm}
 \epsfig{figure=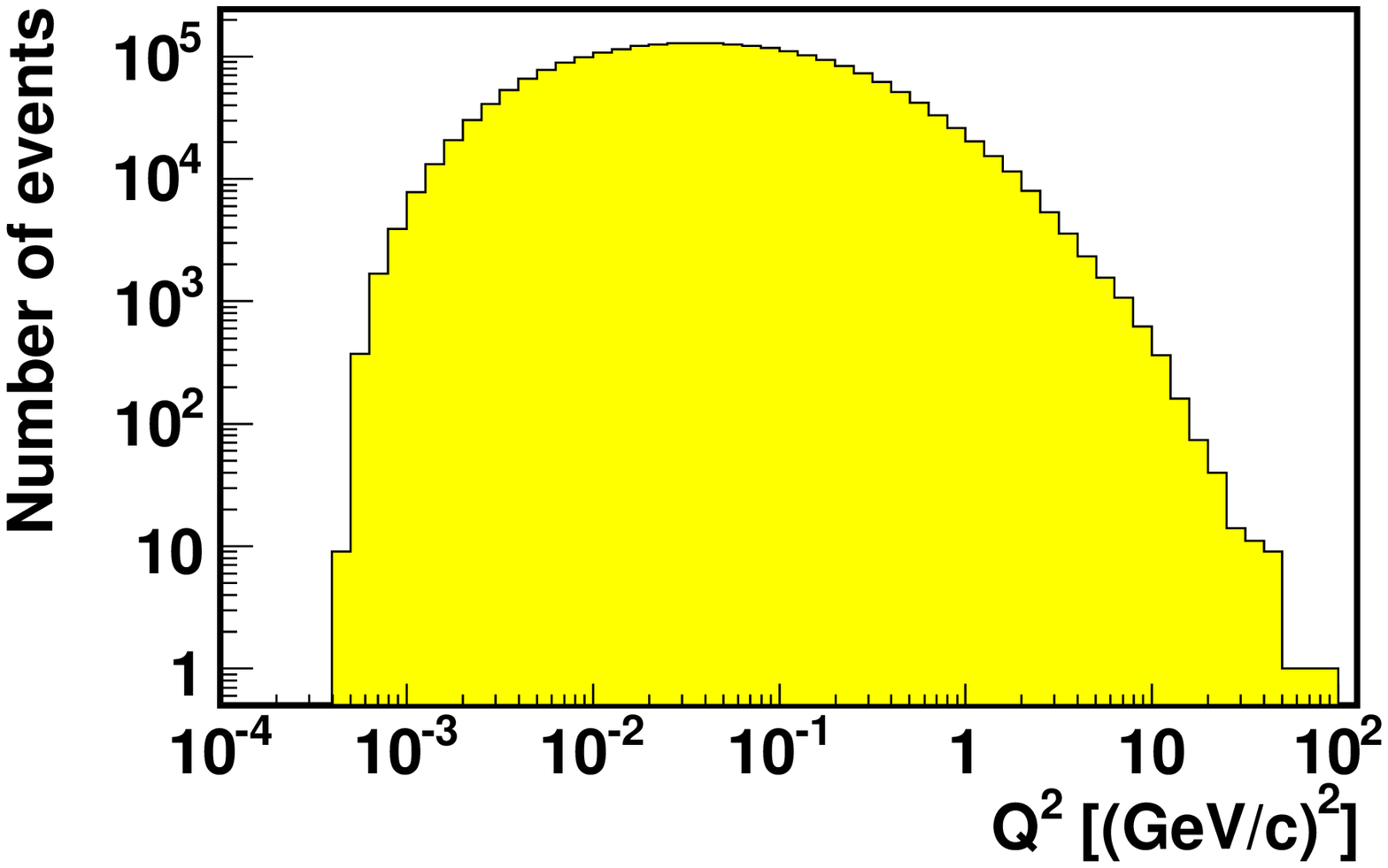,height=5cm,width=7.5cm}
 \epsfig{figure=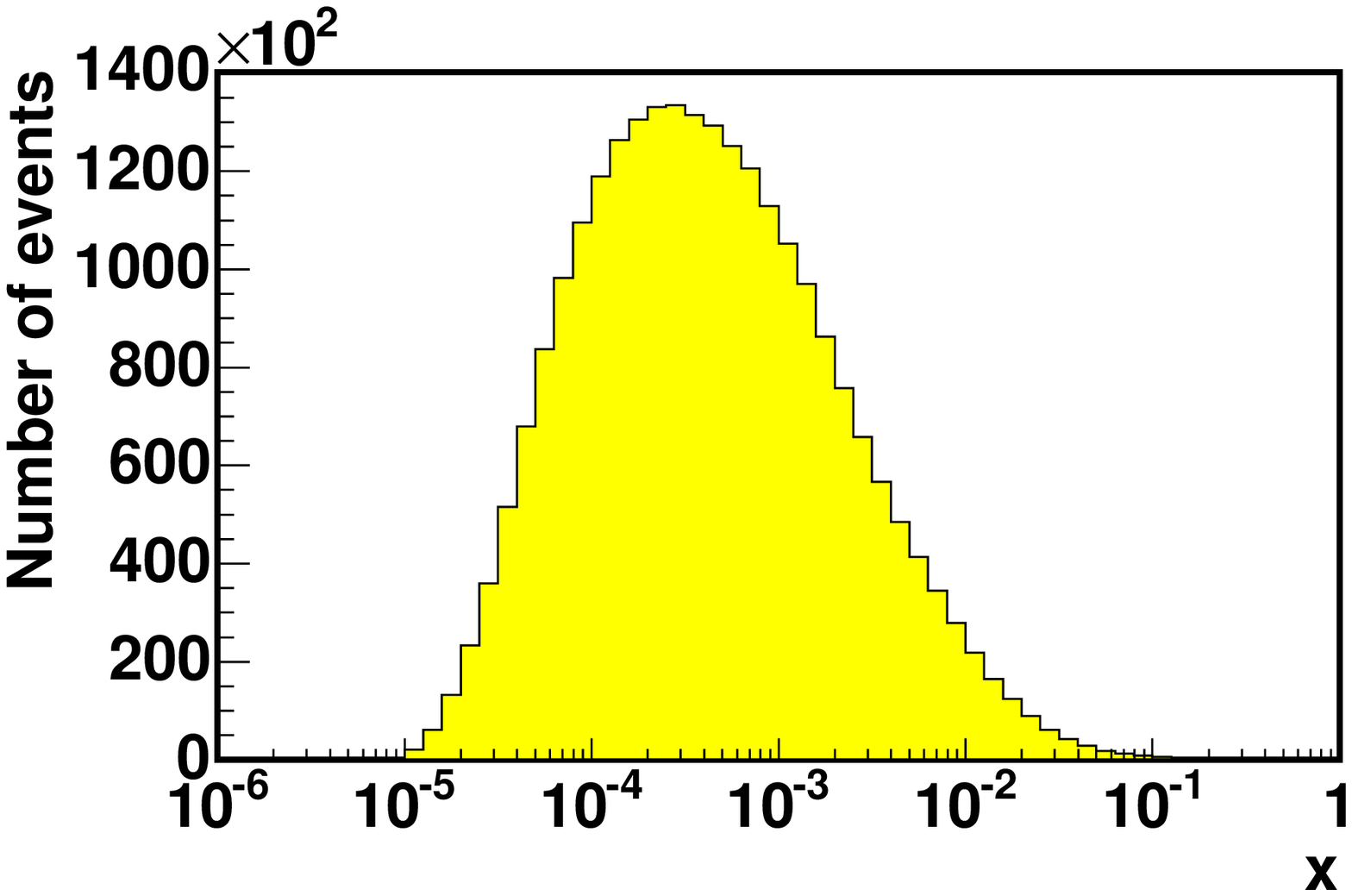,height=5cm,width=7.5cm}
 \epsfig{figure=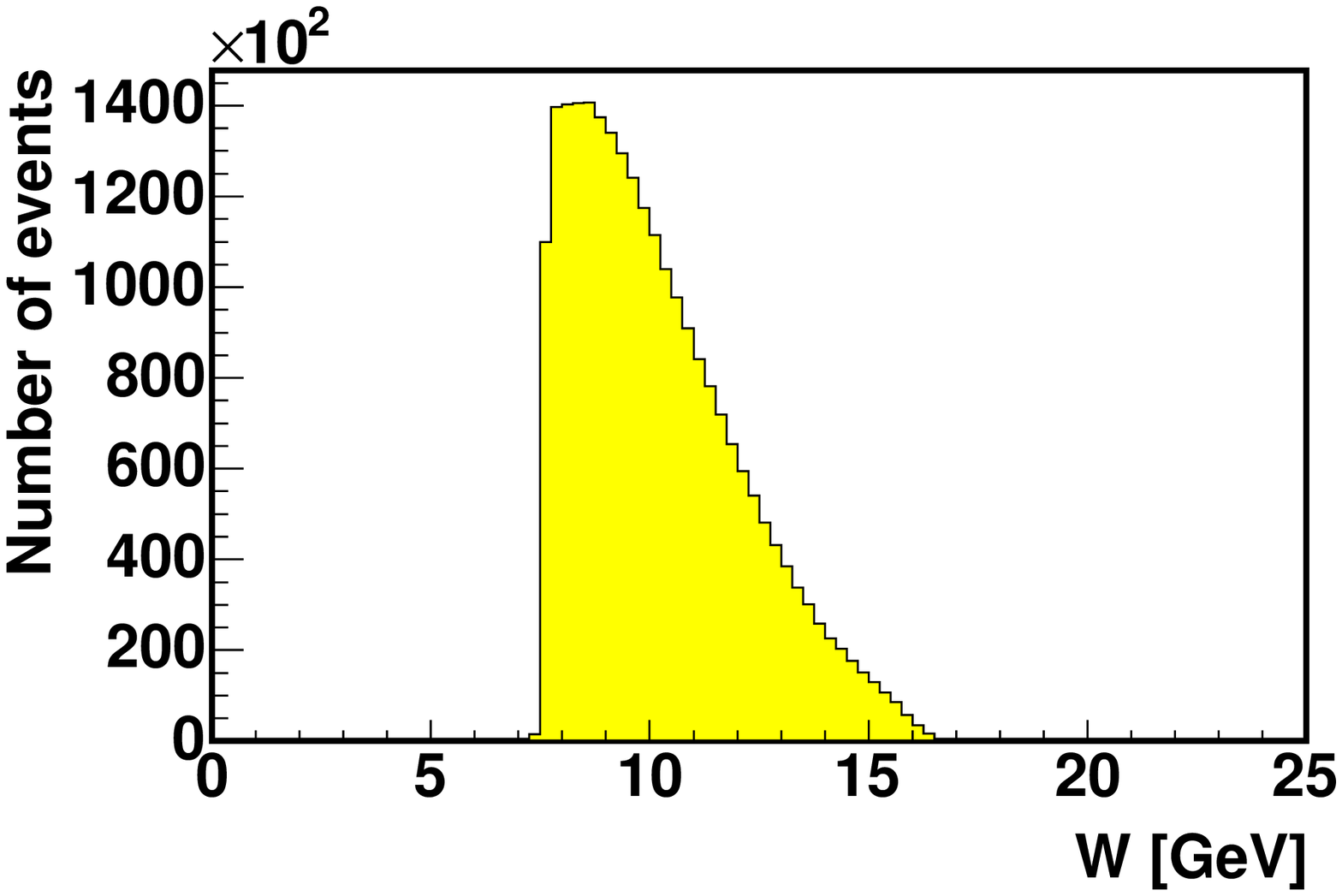,height=5cm,width=7.5cm}
 
 \caption{Distributions of the kinematical variables for the final sample: \Qs\ with linear and
           logarithmic vertical axis scale (top left and right panels respectively), \xBj\ (bottom left), and the energy
           $W$ (bottom right).}
 \label{kinplots}
 \end{center}
\end{figure}

After all selections the final sample consists of about 2.44 million events.
The distributions of \Qs, \xBj\ and $W$ are shown in Fig.~\ref{kinplots}.
The data cover a wide range in \Qs\ and \xBj\ which extends towards the small values 
by almost two orders of magnitude compared to the similar studies reported in 
Ref.~\cite{HERMES03}. The sharp edge of the $W$ distribution at the low $W$ values
is a consequence of the cut applied on $\nu $. For this sample 
$\langle W \rangle$ is equal to 10.2~GeV and $\langle p_{t}^{2}
\rangle = 0.27(\GeV/c)^2$.

\section{Extraction of asymmetry \Aor}
 \label{Sec_extract} 

The cross section asymmetry $A_{LL}  =  
(\sigma_{\uparrow \downarrow}  -  \sigma_{\uparrow \uparrow})/
(\sigma_{\uparrow \downarrow}  +  \sigma_{\uparrow \uparrow})$ for reaction
(\ref{murho}) , for antiparallel ($\uparrow \downarrow $) and parallel 
($\uparrow \uparrow $) spins of
the incoming muon and the target nucleon, is related to the virtual-photon
nucleon asymmetry $A_{1}^{\rho}$ by
\begin{equation}
 A_{LL}  =  D  \left( A_{1}^{\rho}  +  \eta  A_{2}^{\rho}  \right) ,      \label{A1rALL}
\end{equation}
where the factors $D$ and $\eta $ depend on the event kinematics and 
$A_{2}^{\rho}$ is related to
the interference cross section for exclusive production by 
longitudinal and transverse virtual photons. 
As the presented results extend into the range of very small $Q^2$, 
the exact formulae for the depolarisation factor $D$ and kinematical 
factor $\eta$ \cite{JKir}
are used without neglecting terms proportional to the lepton mass squared $m^2$.
The depolarisation factor is given by
\begin{equation}
 D(y, Q^{2}) = 
   \frac{
    y \left[ (1 + \gamma^{2} y/2) (2  -  y) - 
    2 y^{2} m^{2} / Q^{2}  \right]
   }{
    y^{2}  (1  -  2 m^{2} / Q^{2})  (1  +  \gamma^{2})  + 
    2  (1  +  R)  (1  -  y  -  \gamma^{2} y^{2}/4)
   },    \label{depf}
\end{equation}
where $R  =  \sigma_{L} / \sigma_{T}$, $\sigma_{L(T)}$ is the cross section for reaction (\ref{phorho})
initiated by longitudinally (transversely) polarised virtual photons, the fraction of the muon energy lost $y = \nu /E_{\mu }$ and $\gamma^{2}  =  Q^{2} / \nu^{2}$.
The kinematical factor $\eta (y, Q^{2})$  is the same as for the inclusive 
asymmetry.

\begin{figure}[thb]
\begin{center}
  \epsfig{file=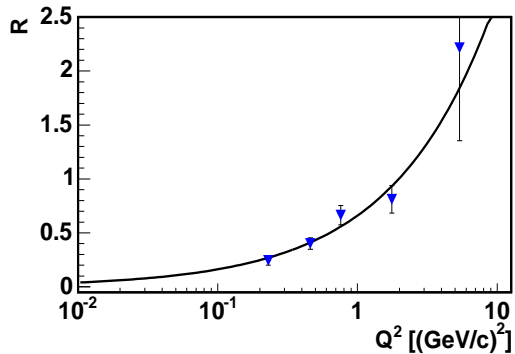,height=5cm,width=7.4cm}
  \caption{The ratio $R= \sigma_L/\sigma_T$ as a function of \Qs\ measured in the E665 experiment.
           The curve is a fit to the data described in the text.}
  \label{R-pict}
 \end{center}
\end{figure}

The asymmetry $A_{2}^{\rho}$  obeys the positivity limit $A_{2}^{\rho} < \sqrt{R}$, analogous to the one for the inclusive case.
For $Q^2 \leq  0.1~(\GeV/c)^2$ the ratio $R$ for the reaction (\ref{phorho}) is
small, cf.\ Fig.~\ref{R-pict}, and the positivity limit constrains $A_{2}^{\rho}$ to small values. Although for larger \Qs\ the ratio $R$ for the process (\ref{phorho}) increases
with \Qs, because of small values of $\eta $ the product $\eta \sqrt{R}$ is small in the  whole \Qs\ range of our sample.
Therefore the second term in Eq.~\ref{A1rALL} can be neglected, so that
\begin{equation}
 A_{1}^{\rho} \simeq \frac{1}{D}  A_{LL},
\end{equation}
and the effect of this approximation is included in the systematic uncertainty of \Aor.

The number of events $N_i$ collected from a given target cell in a given time interval
is related to the spin-independent cross section $\bar{\sigma}$ for reaction (\ref{phorho})
and to the asymmetry $A_1^{\rho}$ by
\begin{equation}
N_i =  a_i \phi _i n_i \bar{\sigma } (1+P_B P_T f D A_1^{\rho }  ),
\label{nevents}
\end{equation}
where $P_B$ and $P_T$ are the beam and target polarisations, $\phi _i$ is the incoming
muon flux, $a_i$ the acceptance for the target cell, $n_i$ corresponding number of target nucleons, and $f$ the target dilution factor.
The asymmetry is extracted from the data sets taken before and after a reversal of the
target spin directions. The four relations of Eq.~\ref{nevents}, corresponding to the two 
cells ($u$ and $d$) and the two spin orientations (1 and 2) lead to a second-order
equation in \Aor\ for the ratio $(N_{u,1}N_{d,2}/N_{d,1}N_{u,2})$. Here fluxes cancel out
as well as acceptances, if the ratio of acceptances for the two cells is the same before
and after the reversal \cite{SMClong}. In order to minimise the statistical error
all quantities used in the asymmetry calculation are evaluated event by event with the
weight factor $w=P_BfD$. The polarisation of the beam muon, $P_B$, is obtained from
a simulation of the beam line and parameterised as a function of the beam momentum. The target polarisation is not 
included in the event weight $w$ because it may vary in time and generate false
asymmetries. An average $P_T$ is used for each target cell and each spin orientation.

 The ratio $R$, which enters the formula for $D$ and strongly depends on \Qs\ for reaction (\ref{phorho}), was calculated on an event-by-event basis
using the parameterisation 
\begin{equation}
 R(Q^{2})  = 
   a_{0}  (Q^{2})^{a_{1}} ,
\end{equation}
with $a_{0} = 0.66 \pm 0.05$, and $a_{1} = 0.61 \pm 0.09$.
The parameterisation was obtained by the Fermilab E665 experiment from a fit
to their $R$ measurements for exclusive $\rho ^0$ muoproduction on protons \cite{E665-1}.
These are shown in Fig.~\ref{R-pict} together with the fitted $Q^2$-dependence.
The preliminary COMPASS results on $R$ for the incoherent exclusive $\rho ^0$
production on the nucleon \cite{sandacz}, which cover a broader kinematic region in $Q^2$
, agree reasonably well with this parameterisation.
The uncertainty of $a_{0}$ and $a_{1}$ is included in the 
systematic error of $A_{1}^{\rho}$.

The dilution factor $f$ gives the fraction of events of reaction (\ref{phorho}) 
originating from nucleons in polarised deuterons inside the target material.
It is calculated event-by-event using the formula
\begin{equation}
 f  =  C_1  \cdot f_{0}  =  C_1  \cdot 
   \frac{
    n_{\rm D}
   }{
    n_{\rm D} +  \Sigma_{A}  n_{\rm A} 
    (\tilde{\sigma}_{\rm A} / \tilde{\sigma}_{\rm D})
   }.    
   \vspace*{3mm}
\end{equation}
Here $n_{\rm D}$ and $n_{\rm A}$ denote numbers of nucleons in deuteron and nucleus of atomic mass $A$
in the target, and 
$\tilde{\sigma}_{\rm D}$ and $\tilde{\sigma}_{\rm A}$ are the cross sections
 per nucleon for reaction (\ref{phorho}) occurring on the deuteron and on the nucleus of atomic mass
$A$, respectively. The sum runs over all nuclei present in the COMPASS target.
The factor $C_1$ takes into account that there are two polarised deuterons in
the $^6$LiD molecule,
as the $^6$Li nucleus is in a first approximation composed of a deuteron and
an $\alpha $ particle.

\begin{figure}[t]
\begin{center}
  \epsfig{file=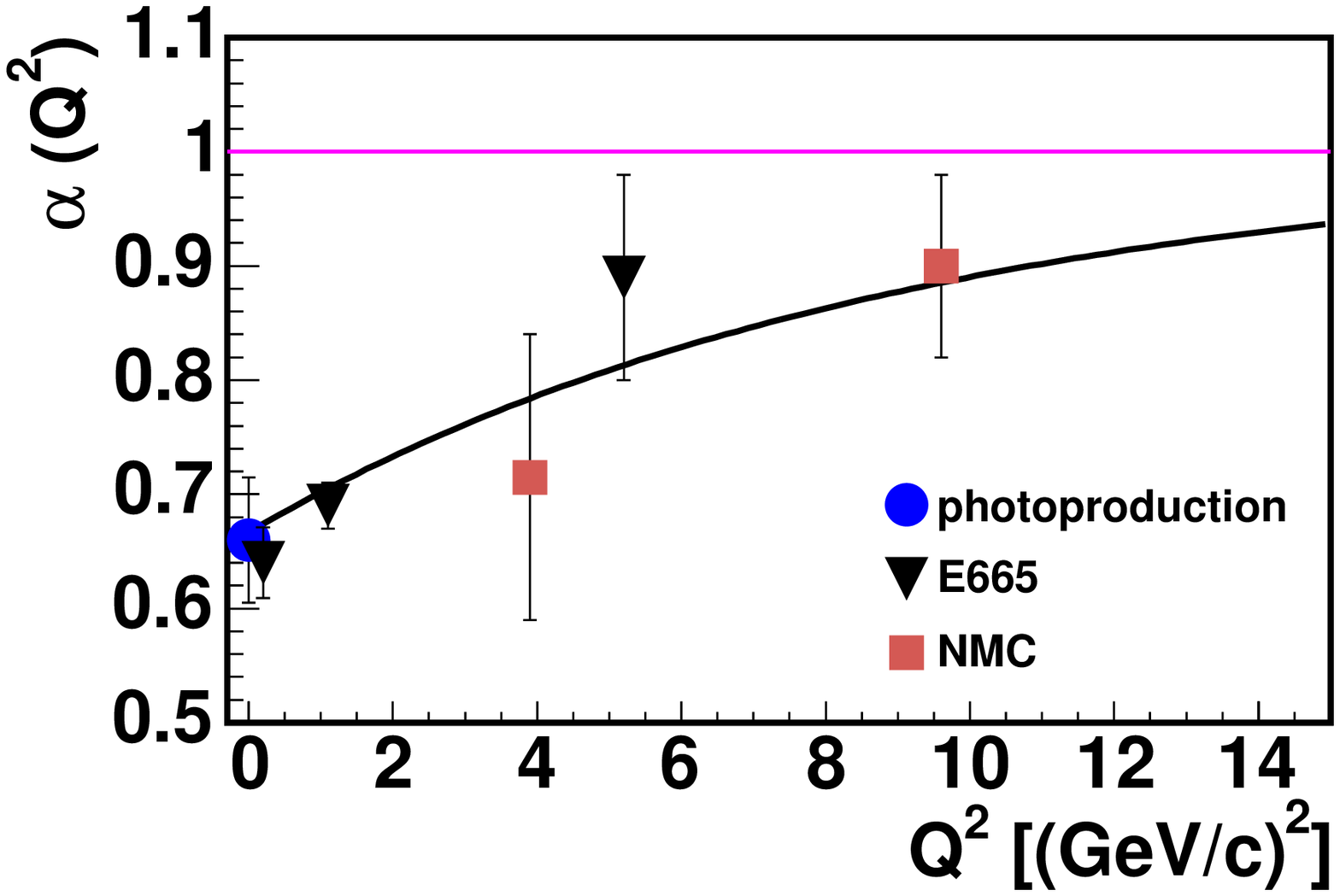,height=5cm,width=7.5cm}
  \epsfig{file=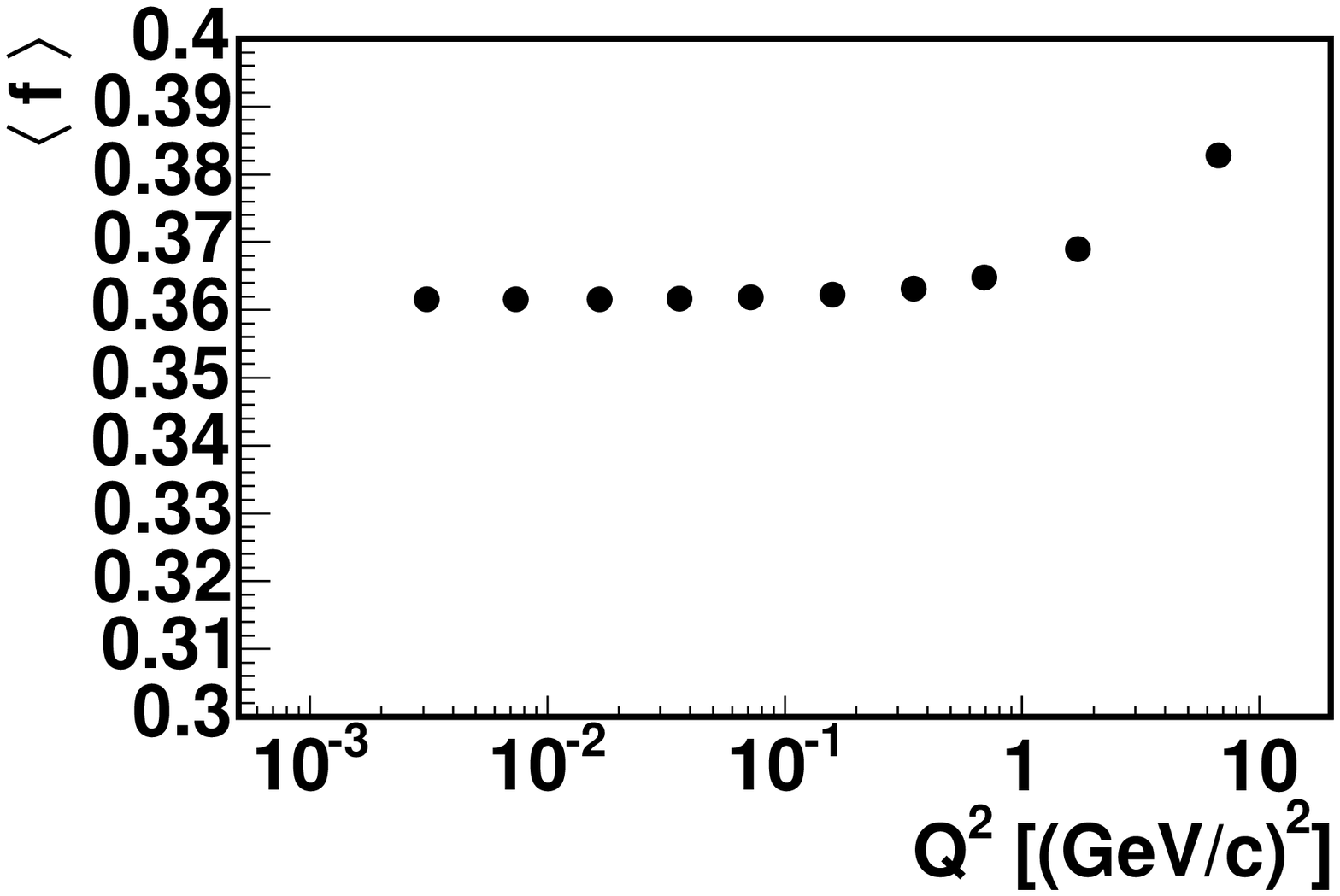,height=5cm,width=7.5cm}
  \caption{(Left) Parameter $\alpha$ of Eq.~\ref{alpha} as a function of \Qs\ (from Ref.~\cite{ATrip-2}). The experimental points and the fitted curve are shown. See text for details. (Right) The dilution factor $f$ as a function of \Qs.}
  \label{alpha-fig}
 \end{center}
\end{figure}

The measurements of the $\tilde{\sigma}_{\rm A} / \tilde{\sigma}_{\rm D}$ 
for incoherent exclusive $\rho ^0$ production come from the NMC \cite{NMC}, E665 \cite{E665-2} and early experiments on \rn\ photoproduction \cite{BPSY}. They were fitted in Ref.~\cite{ATrip-2}
with the formula:
\begin{equation}
 \tilde{\sigma}_{\rm A}  = 
   \sigma_{\rm p} \cdot A^{\alpha(Q^{2})  -  1} , \hspace*{1cm}
    {\rm with}\ \ \alpha(Q^{2})  -  1  =  -\frac{1}{3}  \exp\{-Q^{2}/Q^{2}_{0}\},
    \label{alpha}
\end{equation}
where $\sigma_{\rm p}$ is the cross section for reaction (\ref{phorho})
on the free proton.
The value of the fitted parameter $Q^{2}_{0}$ is  equal to $9 \pm 3~(\GeV/c)^2$.
The measured values of the parameter $\alpha $ and the fitted curve $\alpha (Q^2)$
are shown on the left panel of 
Fig.~\ref{alpha-fig} taken from Ref.~\cite{ATrip-2}.
On the right panel of the figure the average value of $f$ is plotted for
the various $Q^2$ bins used in the present analysis. The values of $f$
are equal to about 0.36 in most of the $Q^2$ range, rising to about 0.38 at the
highest $Q^2$.

  The radiative corrections (RC) have been neglected in the present analysis, in particular in the calculation of $f$, because
they are expected to be small for reaction (\ref{murho}). They were evaluated \cite{KKurek} to be
of the order of 6\% for the NMC exclusive \rn\ production analysis.
The small values of RC are mainly due to the requirement of event exclusivity via cuts on \Emi\ and $p_t^2$,
which largely suppress the dominant external photon radiation.
The internal (infrared and virtual) RC were
estimated in Ref.~\cite{KKurek} to be of the order of 2\%.  

\section{Systematic errors}
 \label{Sec_syst}

The main systematic uncertainty of \Aor\ comes from an estimate of possible false asymmetries. 
In order to improve the accuracy of this estimate,
in addition to the standard sample of incoherent events, a second sample
was selected by changing the $p_t^2$ cuts to
\begin{equation}
 0 < p_{t}^{2} < 0.5~(\GeV/c)^{2},    \label{ptsext}
\end{equation}  and keeping all the remaining selections and cuts the same
as for the `incoherent sample'.
In the following it will be
 referred to as the `extended $p_t^2$ sample'.
Such an extension of the \pts\ range allows one to obtain a sample which is
about
five times larger than the incoherent sample. 
However, in addition to incoherent events such a sample contains a large fraction of events originating from coherent \rn\ production.
Therefore, for the estimate of the dilution factor $f$ a different nuclear dependence of the
exclusive cross section was used, applicable for the sum of coherent and incoherent
cross sections \cite{NMC}.
The physics asymmetries \Aor\ for both samples are consistent within statistical errors.

Possible, false experimental asymmetries were searched for by modifying the selection
of data sets used for the asymmetry calculation. The grouping of the data into
configurations with opposite target-polarisation was varied from large samples,
covering at most two weeks of data taking, into about 100 small samples, taken in time
intervals of the order of 16 hours.
A statistical test was performed on the distributions of asymmetries obtained
from these small samples. In each of the \Qs\ and \xBj\ bins the dispersion of the values of
$A_{1}^{\rho}$ around their mean agrees with the statistical error.
Time-dependent effects which would lead to a broadening of these distributions
were thus not observed. Allowing the dispersion of $A_{1}^{\rho}$ to vary within its
two standard deviations we 
obtain for each bin an upper bound for the systematic error arising from
time-dependent effects
\begin{equation}
 \sigma_{\rm falseA,tdep}  < 
   0.56~\sigma_{\rm stat}.
\end{equation}
Here $\sigma _{\rm stat}$ is the statistical error on $A_{1}^{\rho}$ for the extended
$p_t^2$ sample.
The uncertainty on the estimates of possible false asymmetries due to the time-dependent 
effects is the dominant contribution to the total systematic error in most of the kinematical
region.

Asymmetries for configurations where spin effects cancel out were calculated to
check the cancellation of effects due to fluxes and acceptances. They were found compatible
with zero within the statistical errors. Asymmetries obtained with different
settings of the microwave (MW) frequency, used for DNP, were compared in order to test possible
effects related to the orientation of the target magnetic field. 
The results for the extended $p_t^2$ sample tend to show that there is a small
difference between asymmetries for the two MW configurations.
However, because the numbers of events of the data samples taken with each MW setting 
are approximately balanced, the effect of this difference on \Aor\ is
negligible for the total sample.
 
The systematic error on $A_{1}^{\rho}$ also contains an overall scale uncertainty of
6.5\% due to uncertainties on $P_B$ and $P_T$. The uncertainty of the 
parameterisation of $R(Q^2)$ 
affects the depolarisation factor $D$. The uncertainty of the dilution factor
$f$ is mostly due to uncertainty of the parameter $\alpha (Q^2)$ which takes into account
nuclear effects in the incoherent $\rho ^0$ production. 
The neglect of the $A_2^{\rho }$ term mainly affects the highest bins of \Qs\ and \xBj.

Another source of systematic errors is due to the contribution of
the non-exclusive background to our sample.
This background originates from two sources.
First one is due to the  production of
\rn\ accompanied by the dissociation of the target nucleon,
the second one is the production of \rn\ in inclusive scattering.
In order to evaluate the amount of background in the sample of exclusive
events it is necessary to
determine the $E_{miss}$ dependence for the non-exclusive background
in the region under the exclusive peak (cf.\ Fig.~\ref{hadplots} ).
For this purpose complete Monte Carlo simulations of the experiment were 
used, with events 
generated by either the PYTHIA 6.2 or LEPTO 6.5.1 generators.
Events generated with LEPTO come only from deep inelastic scattering and cover the range of $Q^2 > 0.5~(\GeV/c)^2$. 
Those generated with PYTHIA cover the whole kinematical range of the experiment and include exclusive production of vector mesons and processes with diffractive excitation of the target nucleon or the vector meson, in addition to inelastic
production.

The generated MC events were reconstructed
and selected for the analysis using the same procedure as for the data.
In each bin of \Qs\ the $E_{miss}$ distribution for the MC was 
normalised to the corresponding one for the data
in the range of large $E_{miss} > 7.5~\GeV$. Then the normalised MC
distribution was used to estimate the number of background events under the exclusive peak in the data.  
The fraction of background events in the sample of
incoherent exclusive $\rho ^0$ production was estimated to be about $0.12 \pm 0.06$
in most of the kinematical range, except in the largest \Qs\ region, where it is about
$0.24 \pm 0.12$. 
The large uncertainties of these fractions reflect the differences between 
estimates from LEPTO and PYTHIA in the region where they overlap. In 
the case of PYTHIA the
uncertainties on the cross sections for diffractive photo- and
electroproduction of vector mesons also contribute.
For events generated with PYTHIA the $E_{miss}$  
distributions for various physics processes could be studied separately. 
It was found
that events of \rn\ production with an
excitation of the target nucleon into $N^*$ resonances
of small mass, $M < 2~\GeV/c^2$, cannot be resolved from the exclusive
peak and therefore were not included                 
in the estimates of number of background events.

An estimate of the asymmetry \Aor\ for the background was obtained using a non-exclusive sample,
which was selected with the standard cuts used in this analysis,
except the cut on $E_{miss}$ which was modified to $E_{miss} > 2.5$~GeV. In different high-$E_{miss}$ bins \Aor\ for this sample was
found compatible with zero.

Because no indication of a non-zero \Aor\ for the background was found,
and also due to a large uncertainty of the estimated amount of background 
in the exclusive sample, no background corrections were made.
Instead, the effect of background was treated as a source
of systematic error. Its contribution 
to the total systematic error was not significant in most of the kinematical
range, except for the highest $Q^2$ and $x$.

The total systematic error on \Aor\ was obtained as a quadratic sum of the errors from
all discussed sources. 
Its values for each \Qs\ and \xBj\ bin are given
in Tables~\ref{binsq2-1}  and \ref{binsx-1}. The total systematic error 
amounts to about 40\% of the 
statistical error for most of the kinematical range. Both errors become
comparable in the highest bin of \Qs. 

\section{Results}
 \label{Sec_resu}

\begin{figure}[ht]
\begin{center}
  \epsfig{file=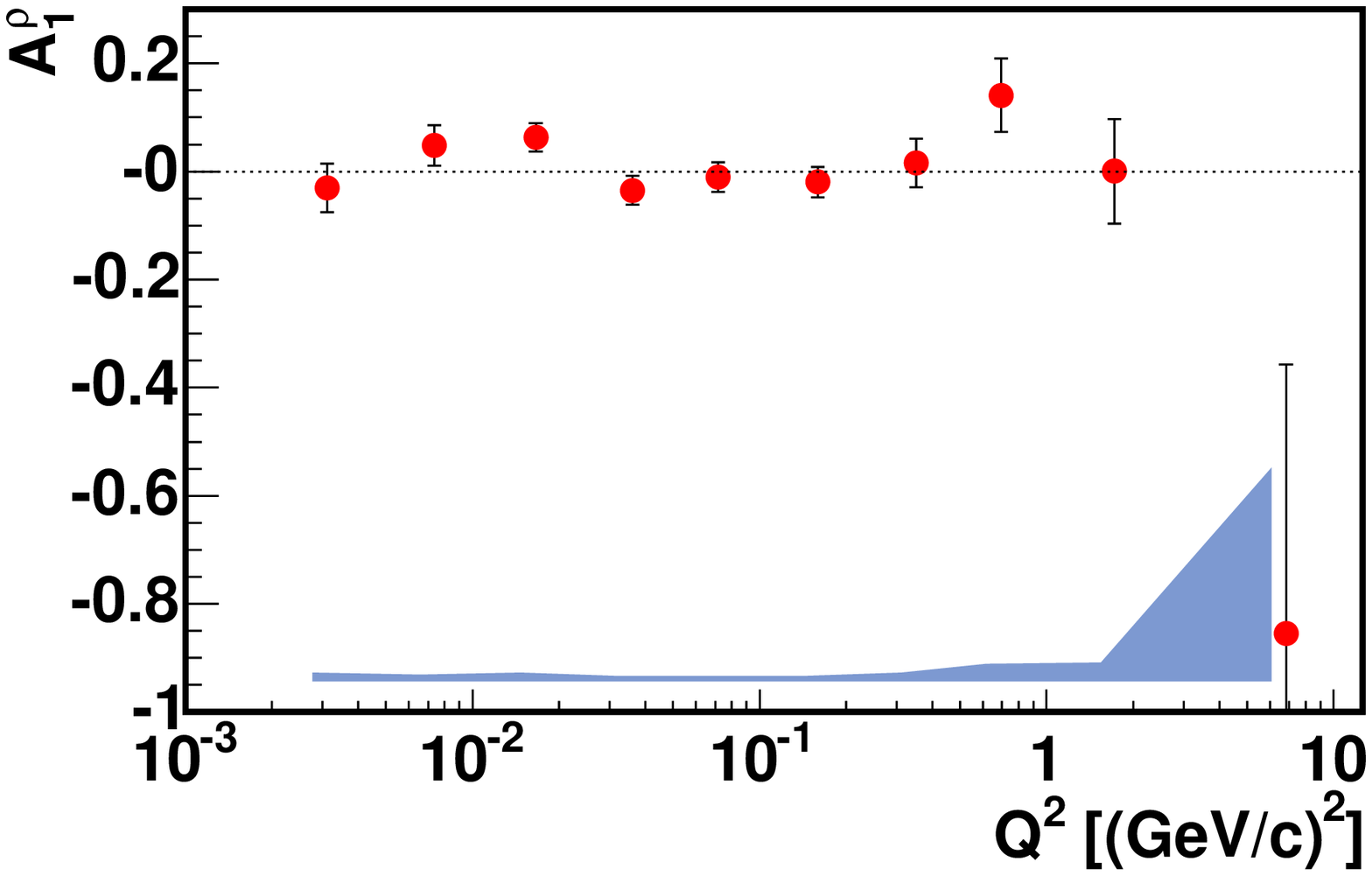,height=5cm,width=7.5cm}
  \epsfig{file=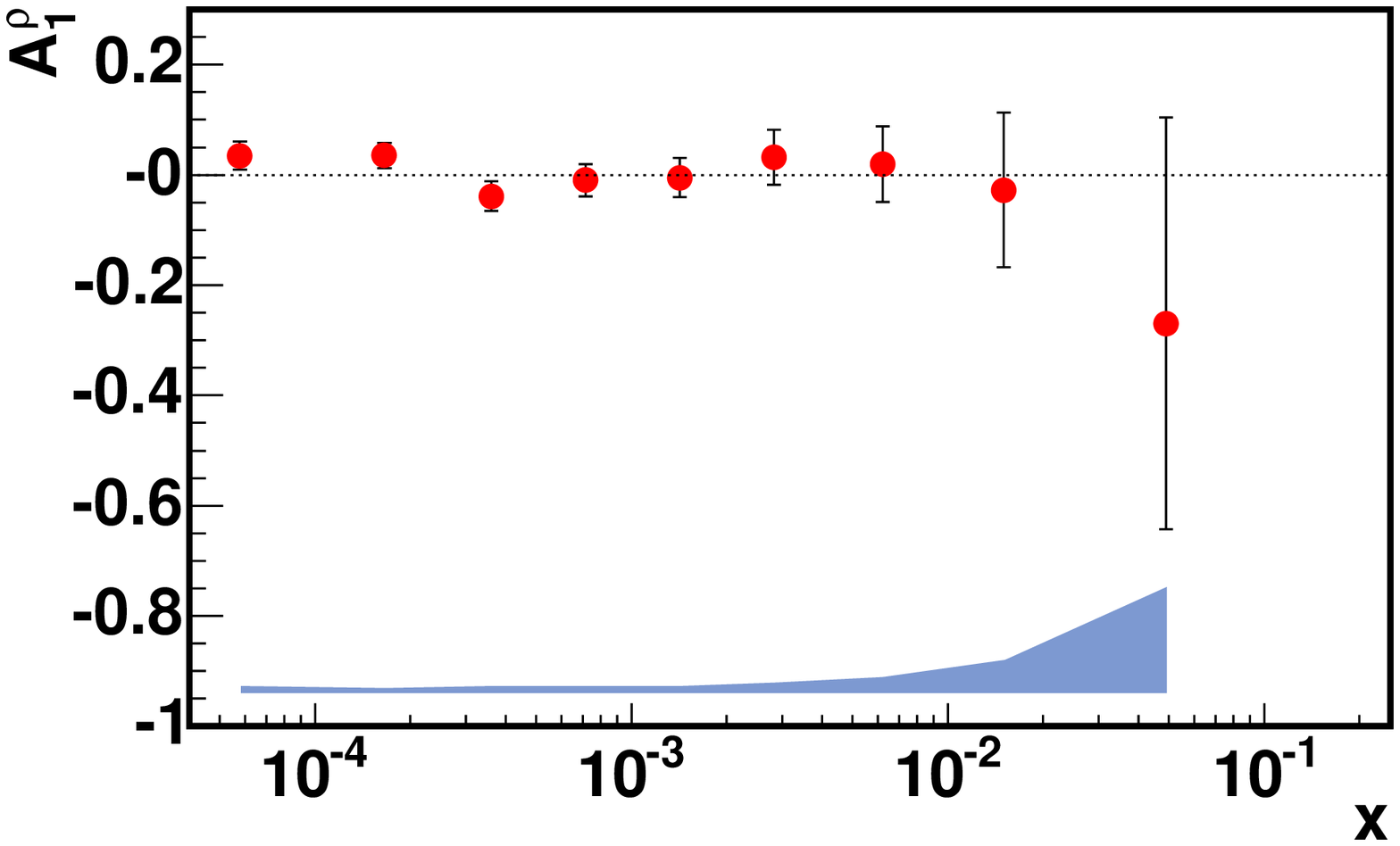,height=5cm,width=7.5cm}
  \caption{\Aor\ as a function of \Qs\ (left) and \xBj\ (right) from the present analysis.
Error bars correspond to statistical errors, while bands at the bottom represent the systematical errors.}
  \label{A1-Com}
 \end{center}
\end{figure}

The COMPASS results on $A^{\rho}_1$ are shown as a function of \Qs\  and \xBj\ 
in Fig.~\ref{A1-Com} and listed in 
Tables~\ref{binsq2-1} and \ref{binsx-1}.
The statistical errors are represented by vertical bars and the total systematic
errors by shaded bands.

\begin{table}[ht]
 \caption{Asymmetry $A_1^{\rho }$ as a function of \Qs. Both the statistical errors (first) and the total systematic errors (second) are listed.}
 \label{binsq2-1}
 \small
 \begin{center}
  \begin{tabular}{||c|c|c|c|c||}
   \hline
   \hline
   \raisebox{0mm}[4mm][2mm]{\Qs\ range}\ \  &
                            $\langle  Q^{2}  \rangle$ [$(\GeV/c)^2$]  &  $\langle  x  \rangle$  &
                            $\langle  \nu  \rangle$ [GeV] & $A_1^{\rho }$ \\   \hline
   \hline
   \raisebox{0mm}[4mm][2mm] \ \ $0.0004 - 0.005~$  & \ \ \ 0.0031\ \ \  & $4.0 \cdot 10^{-5}$ & 42.8  & $-0.030 \pm 0.045 \pm 0.014$\\
   \hline
   \raisebox{0mm}[4mm][2mm] \ \ $0.005 - 0.010$  &  ~0.0074 & $8.4 \cdot 10^{-5}$ & 49.9             & $~~0.048 \pm 0.038 \pm 0.013$\\
   \hline
   \raisebox{0mm}[4mm][2mm] \ \ $0.010  - 0.025$ &  0.017  & $1.8 \cdot 10^{-4}$ & 55.6             & $~~0.063 \pm 0.026 \pm 0.014$\\
   \hline
   \raisebox{0mm}[4mm][2mm] \ \ $0.025 - 0.050$  &  0.036  & $3.7 \cdot 10^{-4}$ & 59.9             & $-0.035 \pm 0.027 \pm 0.009$\\
   \hline
   \raisebox{0mm}[4mm][2mm] \ \ $0.05  - 0.10$   &  0.072  & $7.1 \cdot 10^{-4}$ & 62.0             & $-0.010 \pm 0.028 \pm 0.008$\\
   \hline
   \raisebox{0mm}[4mm][2mm] \ \ $0.10   - 0.25$  &  0.16~  & 0.0016 & 62.3                           & $-0.019 \pm 0.029 \pm 0.009$\\
   \hline
   \raisebox{0mm}[4mm][2mm] \ \ $0.25  - 0.50$   &  0.35~  & 0.0036 & 60.3                           & $~~0.016 \pm 0.045 \pm 0.014$\\
   \hline
   \raisebox{0mm}[4mm][2mm] \ \ $0.5   - 1~~$     &  0.69~  & 0.0074 & 58.6                           & $~~0.141 \pm 0.069 \pm 0.030$\\
   \hline
   \raisebox{0mm}[4mm][2mm] \ \ $1     - 4$     &  1.7~~  & 0.018~~  & 59.7                           & $~~0.000 \pm 0.098 \pm 0.035$\\
   \hline
   \raisebox{0mm}[4mm][2mm] \ \ $~4    - 50$  & 6.8~~  & 0.075~~  & 55.9                               & $-0.85  \pm 0.50  \pm 0.39 $\\
   \hline
   \hline
  \end{tabular}
 \end{center}
  \normalsize
\end{table}
\begin{table}[ht]
 \caption{Asymmetry $A_1^{\rho }$ as a function of \xBj. Both the statistical errors (first) and the total systematic errors (second) are listed.}
 \label{binsx-1}
 \small
 \begin{center}
  \begin{tabular}{||c|c|c|c|c||}
   \hline
   \hline
   \raisebox{0mm}[4mm][2mm]{\xBj\ range}\ \  &   $\langle  x  \rangle$  &  $\langle  Q^{2}  \rangle$ [$(\GeV/c)^2$]  &
                           $\langle  \nu  \rangle$ [GeV] & $A_1^{\rho }$  \\   \hline
   \hline
   \raisebox{0mm}[4mm][2mm] \ \ $8 \cdot 10^{-6} - 1 \cdot 10^{-4}$   & $5.8 \cdot 10^{-5}$ & ~0.0058 & 51.7 & $~~0.035 \pm 0.026 \pm 0.011$ \\
   \hline
   \raisebox{0mm}[4mm][2mm] \ \ $1 \cdot 10^{-4} - 2.5 \cdot 10^{-4}$  & $1.7 \cdot 10^{-4}$ & 0.019 &  59.7  & $~~0.036 \pm 0.024 \pm 0.010$\\
   \hline
   \raisebox{0mm}[4mm][2mm] \ \ $2.5 \cdot 10^{-4} - 5 \cdot 10^{-4}$   & $3.6 \cdot 10^{-4}$ & 0.041 &  61.3  & $-0.039 \pm 0.027 \pm 0.012$\\
   \hline
   \raisebox{0mm}[4mm][2mm] \ \ $5 \cdot 10^{-4} - 0.001$ &  $7.1 \cdot 10^{-4}$ & 0.082 & 60.8  & $-0.010 \pm 0.030 \pm 0.010$\\
   \hline
   \raisebox{0mm}[4mm][2mm] \ \ $0.001 - 0.002$ &  0.0014 & 0.16~~ & 58.6 & $-0.005 \pm 0.036 \pm 0.013$ \\
   \hline
   \raisebox{0mm}[4mm][2mm] \ \ $0.002 - 0.004$ &  0.0028 & 0.29~~ & 54.8 & $ ~~0.032 \pm 0.050 \pm 0.019$ \\
   \hline
   \raisebox{0mm}[4mm][2mm] \ \ $0.004 - 0.01~~$  &  0.0062 & 0.59~~ & 50.7 & $ ~~0.019 \pm 0.069 \pm 0.026$ \\
   \hline
   \raisebox{0mm}[4mm][2mm] \ \ $0.01  - 0.025$ &  0.015~~  & 1.3~~~~  & 47.5 & $-0.03 \pm 0.14 \pm 0.06$ \\
   \hline
   \raisebox{0mm}[4mm][2mm] \ \ $0.025 - 0.8~~~~$  &   0.049~~  & 3.9~~~~  & 43.8 & $-0.27 \pm 0.38 \pm 0.19$  \\
   \hline
   \hline
  \end{tabular}
 \end{center}
 \normalsize
\end{table}

The wide range in $Q^2$
covers four orders of magnitude from $3 \cdot 10^{-3}$ to 7~$(\GeV/c)^2$. The domain in \xBj\ which is strongly correlated
with $Q^2$, varies from $5 \cdot 10^{-5}$ to about 0.05 (see Tables for more details).
For the whole kinematical range the \Aor\
asymmetry measured by COMPASS is consistent with zero. As discussed in the introduction,
this indicates that the role of unnatural parity exchanges,
like $\pi$- or $A_{1}$-Reggeon exchange, 
is small in that kinematical domain, which is to be expected if diffraction
is the dominant process for reaction (\ref{phorho}). 

\begin{figure}[ht]
\begin{center}
  \epsfig{file=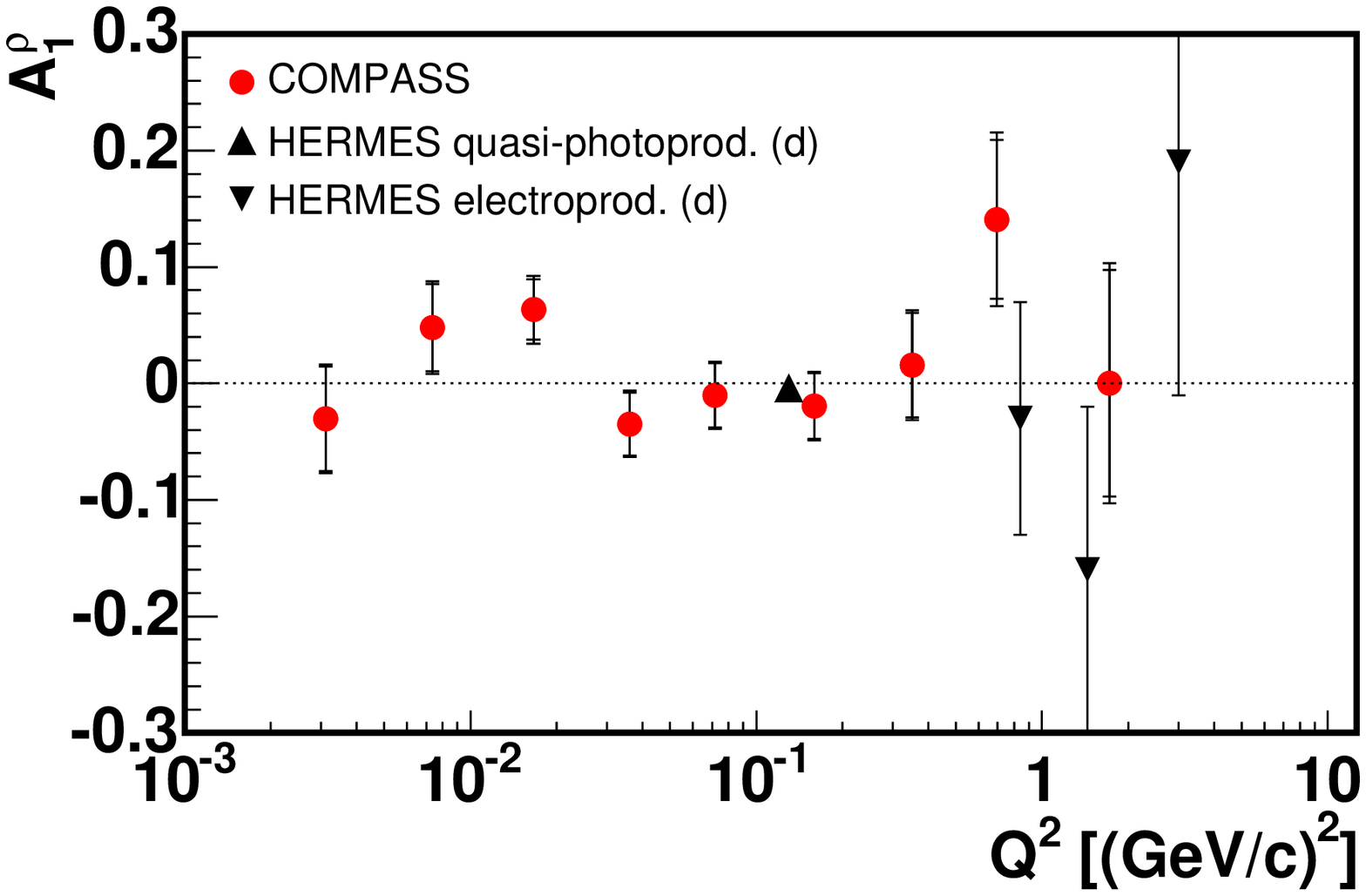,height=5cm,width=7.5cm}
  \epsfig{file=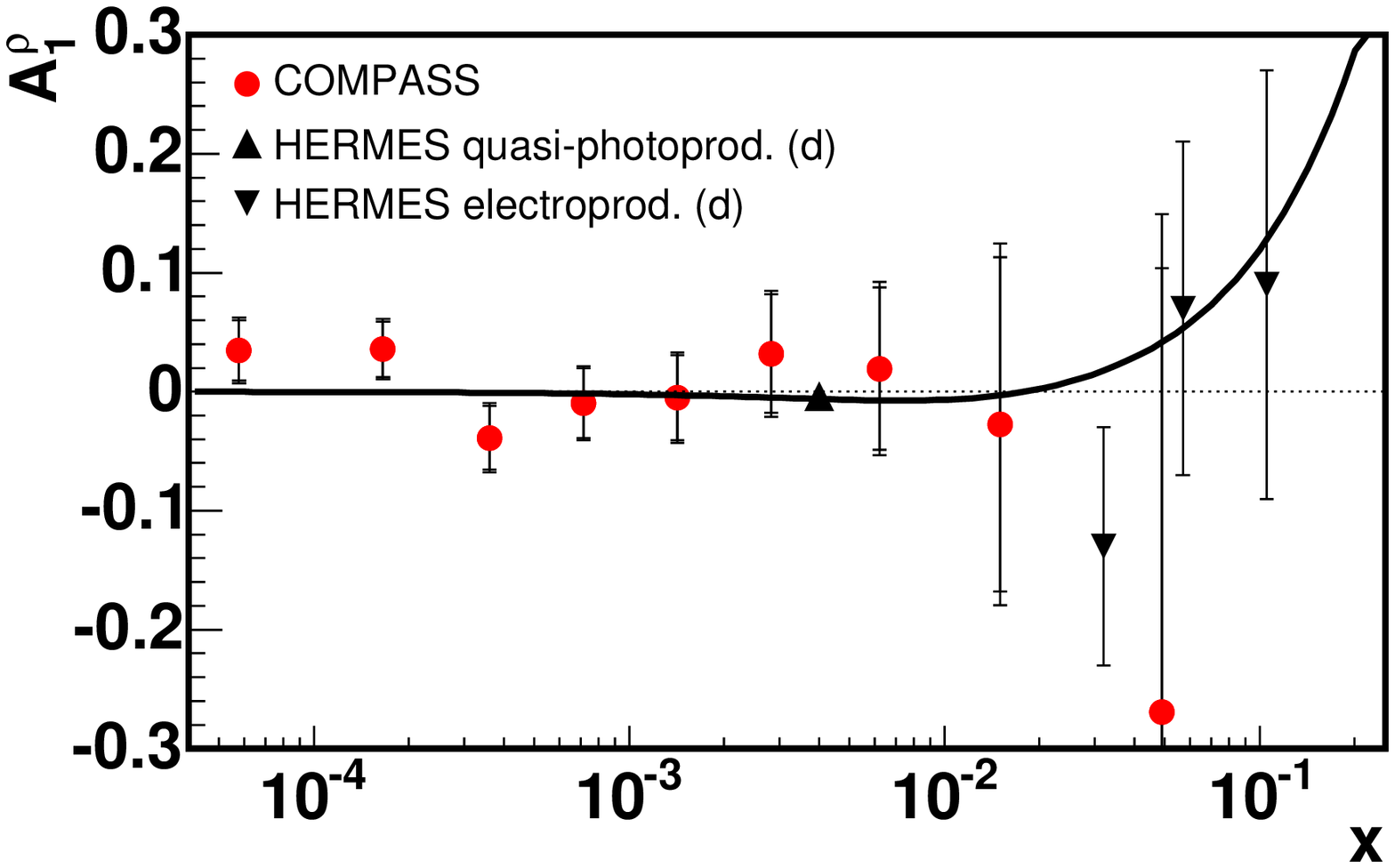,height=5cm,width=7.5cm}
  \caption{\Aor\ as a function of \Qs\ (left) and \xBj\ (right) from the present analysis (circles)
compared to HERMES results on the deuteron target (triangles). For the COMPASS results inner bars represent statistical errors, while the outer bars correspond to the total error.
 For the HERMES results vertical bars represent the quadratic sum of statistical and systematic errors. The curve represents the prediction explained in the text.}
   \label{A1-Com-Her}
\end{center}
\end{figure}

In Fig.~\ref{A1-Com-Her} the COMPASS results are compared to the HERMES results on $A^{\rho}_1$ obtained
on a deuteron target \cite{HERMES03}.
Note that the lowest \Qs\ and \xBj\ HERMES points, referred to as `quasi-photoproduction', come from measurements where the kinematics of the small-angle scattered 
electron was not measured but estimated from a MC simulation. This is in contrast to COMPASS,
where scattered muon kinematics is measured even at the smallest \Qs.

The results from both experiments are consistent within errors. 
The kinematical range covered by the present analysis extends further towards small
values of \xBj\ and \Qs\ by almost two orders of 
magnitude. In each of the two experiments $A^{\rho}_1$ is measured
at different average  $W$, which is equal to about 10~GeV for
COMPASS and 5~GeV for HERMES. Thus, no significant $W$ dependence
is observed for $A^{\rho}_1$ on an isoscalar nucleon target.

The \xBj\ dependence of the measured \Aor\ is compared in Fig.~\ref{A1-Com-Her} to the prediction given by Eq.~\ref{A1-Fraas}, which
relates $A_1^{\rho}$ to the asymmetry $A_{1}$ for the inclusive inelastic
lepton-nucleon scattering. To produce the curve
the inclusive asymmetry $A_{1}$ was parameterised as
$A_1(x) = (x^{\alpha } - \gamma^{\alpha }) \cdot (1- e^{-\beta x})$ , where
$\alpha = 1.158 \pm 0.024$, $\beta = 125.1 \pm 115.7$ and $\gamma = 0.0180 \pm 0.0038$.
The values of the parameters have been obtained
from a fit of $A_{1}(x)$ to the world data
from polarised deuteron targets \cite{smc,e143,e155_d,smc_lowx,hermes_new,compass_a1_recent} including COMPASS measurements at very 
low $Q^2$ and \xBj\ \cite{compass_a1_lowq2}. Within the present accuracy the results on \Aor\ are consistent with this prediction.  


In the highest \Qs\ bin,  $\langle  Q^{2} \rangle = 6.8~(\GeV/c)^2$, in the kinematical domain of applicability of pQCD-inspired models which relate
the asymmetry to the spin-dependent
GPDs for gluons and quarks (cf.\ Introduction), 
one can observe a hint of a possible nonzero asymmetry, although with a
large error.
It should be noted that in Ref.~\cite{ATrip-1} a negative value of $A_{LL}$ different from zero by about 2 standard deviations
was reported at $\langle  Q^{2}  \rangle  = 7.7~(\GeV/c)^2$.
At COMPASS, including the data 
taken with the longitudinally polarised deuteron target in 2004 and 2006 will result in an
increase of
statistics by a factor of about three compared to the present paper,
and thus may help to clarify the issue.
 
For the whole $Q^2$ range future COMPASS data, to be taken with the polarised proton target, would be very valuable for checking if the role of the flavour-blind
exchanges is indeed dominant, as expected for the Pomeron-mediated process.

\section{Summary}
 \label{Sec_summ}

The longitudinal double spin asymmetry \Aor\ for the diffractive muoproduction of \rn\ meson, $\mu + N \rightarrow 
\mu + N + \rho$, has been measured by scattering longitudinally polarised muons off longitudinally polarised deuterons from the
$^6$LiD target and selecting incoherent exclusive $\rho^0$ production.
The presented results for the COMPASS 2002 and 2003 data cover a range of energy $W$ from about 7 to 15~GeV.

The \Qs\ and \xBj\ dependence of \Aor\ is presented in a wide kinematical range
$3 \cdot 10^{-3} \leq Q^{2} \leq 7~(\GeV/c)^2$ and $5 \cdot 10^{-5} \leq x \leq 0.05$.
These results extend the range in \Qs\ and \xBj\ by two orders of magnitude down
with respect to the existing data from HERMES.

The asymmetry $A^{\rho}_1$ is compatible with zero in the whole \xBj\ and
\Qs\ range.
This may indicate that the role of unnatural parity exchanges like $\pi$- or $A_{1}$-Reggeon exchange is
small in that kinematical domain.

The \xBj\ dependence of measured \Aor\ is consistent with the prediction of Ref.~\cite{HERMES01} which relates $A^{\rho}_1$
to the asymmetry $A_{1}$ for the inclusive inelastic lepton--nucleon scattering.

\section{Acknowledgements}
 \label{Sec_ack}

We gratefully acknowledge the support of the CERN management and staff and
the skill and effort of the technicians of our collaborating institutes.
Special thanks are due to V. Anosov and V. Pesaro for their support during the
installation and the running of the experiment. This work was made possible by 
the financial support of our funding agencies.


\end{document}